\documentclass{aa}
\usepackage{graphicx}

\usepackage{natbib}
\bibpunct[]{(}{)}{;}{a}{}{,}

\usepackage[german,english]{babel}
\newcommand{\Porb}{\mbox{$P_\mathrm{orb}$}}
\newcommand{\Lwd}{\mbox{$L_\mathrm{wd}$}}
\newcommand{\Mwd}{\mbox{$M_\mathrm{wd}$}}
\newcommand{\Rwd}{\mbox{$R_\mathrm{wd}$}}
\newcommand{\Twd}{\mbox{$T_\mathrm{wd}$}}
\newcommand{\Lbl}{\mbox{$L_\mathrm{BL}$}}

\begin{document}

\thesaurus{08(02.01.2;08.02.1;08.09.2\,V446\,Her;08.14.2)}

\title{On the occurrence of dwarf nova outbursts in post novae}

\author{M.R. Schreiber\inst{1}
\and
B.T. G\"ansicke\inst{1}
\and
J.K. Cannizzo\inst{2}}
\institute{Universit\"ats-Sternwarte, Geismarlandstr.11, D-37083 G\"ottingen, Germany
\and
NASA/Goddard Space Flight Center, Laboratory for High Energy Astrophysics, Code 662, Greenbelt, MD\,20771}

\date{Received \underline{\hskip2cm} ; accepted \underline{\hskip2cm} }

\authorrunning{Schreiber et al.}

\maketitle

\begin{abstract}
We show that irradiation of the accretion disc by the white dwarf
limits the occurrence of dwarf nova outbursts in post nova accretion
discs.  After the nova explosion, the white dwarf has to cool for up
to $\sim$\,100\,yr --~depending on the orbital period (i.e., disc
size) and the temperature of the white dwarf after the
nova-eruption~-- before the disc can begin producing dwarf nova
outbursts.  During this time the inner disc is maintained in a hot,
ionised state.  Using these calculations, we interpret the long-term
photometric variability of the post nova V446\,Her (Nova Her 1960)
which shows today regular dwarf nova outbursts. As the white dwarf in
V446\,Her continues to cool over the next $\sim$10-20\,yr, we predict
an increase in the amplitude of outbursts and a decrease of the
outburst frequency, because the decreasing irradiation of the
accretion disc should allow an increasing annular extent of the
accretion disc to participate in limit cycle oscillations.


\end{abstract}

\section{Introduction}
Although nova eruptions and dwarf nova outbursts may resemble each
other in some cases\footnote{In fact, a number of large amplitude/low
outburst frequency dwarf novae are found in the lists of nova
remnants, the most famous example being the short-period dwarf nova WZ
Sge \citep{duerbeck87-1}.}, the physical mechanisms behind the
outbursts are entirely different: a nova eruption arises when
hydrogen-rich material accreted onto the surface of a white dwarf
ignites under degenerate conditions (see \citealt{starrfieldetal98-1}
for a review); dwarf novae outbursts are thought to result from
thermal instabilities associated with hydrogen ionisation in an
accretion disc (see \citealt{cannizzo93-1} for a review and
\citealt{ludwigetal94-1} for a detailed parameter study).

It is clear that the class of cataclysmic variables (CVs), close
binaries containing a white dwarf accreting from a less massive
main-sequence companion, provides numerous potential novae progenitors
(pre-novae). In fact, almost all CVs {\em should} suffer nova
eruptions repeatedly during their lives.  To our knowledge, no system
became a nova {\em after} it had been classified as a CV. However, a
large number of nova remnants (post novae) have been found to be CVs
after they attracted attention by erupting (novae may arise, of
course, on a white dwarf accreting hydrogen-rich material in other
environments, e.g. in a symbiotic binary). Among the wide variety of
known CV subtypes, most post novae were found to have rather high mass
transfer rates $\dot M$, and, thus, fall into the class of novalike
variables\footnote{\label{foot-irrad} Note that the published mass
transfer rates were determined generally under the assumption that the
disc luminosity arises from viscous dissipation only.  However,
irradiation by the hot white dwarf contributes significantly to the
observed disc luminosity in young post novae.}
Only in a small number of cases were quasiperiodic brightenings
observed in post novae, identifying those systems to be dwarf novae
\citep{livio89-1}.

In this paper, we analyse the circumstances under which a post nova
will evolve into a dwarf nova following the nova eruption. In the next
two sections we calculate the irradiation of the accretion disc by the
hot white dwarf, and estimate the amount of time required for a post
nova to cool to the point that it may begin exhibiting dwarf nova
outbursts.  We then discuss our results in the context of the post
nova with the best observational coverage
to test  our predictions
--~V446\,Her.

\begin{figure}
\includegraphics[angle=270,width=9cm]{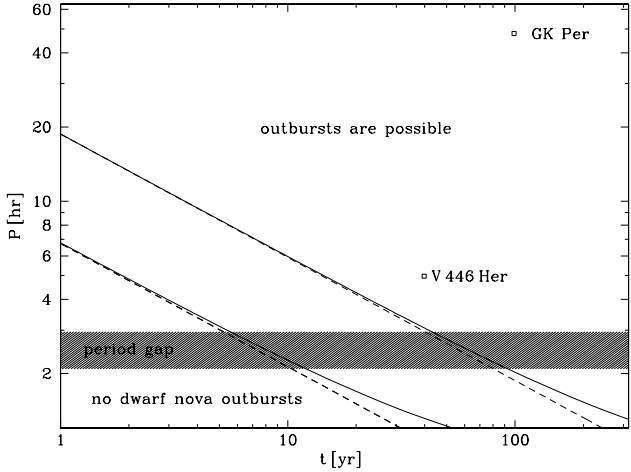}
\caption[]{Irradiation limits on the occurrence of dwarf nova
outbursts as a function of time after nova outburst and orbital
period.  We use initial effective temperatures
$\Twd\,=5\,\times\,10^5\,\mathrm{K}$ (upper curves) and
$\Twd\,=3\,\times\,10^5\,\mathrm{K}$ (lower curves). The solid lines
represent calculations including a boundary layer luminosity
corresponding to $\dot M_\mathrm{wd}=10^{17}\mathrm{g\,s^{-1}}$,
whereas the dashed lines are calculated using only the luminosity of
the cooling white dwarf. Note that $t=0$ corresponds to the end of the
hydrogen shell burning phase, which might last for up to $\sim10$\,yr
after the actual nova eruption.}
\end{figure}

\section{White dwarf cooling in classical novae}
Even though a large part of the accreted hydrogen-rich material is
ejected during the nova eruption, a significant amount of hydrogen is
left over in the remaining envelope, which rapidly returns to
hydrostatic equilibrium \citep{macdonald96-1}. The nova remnant
experiences steady-state hydrogen shell burning until the nuclear fuel
is exhausted, and is consequently heated to effective temperatures of
several $10^5$\,K.  The turn-off times of the burning envelope which
have been observed with ROSAT for a small number of novae are of the
order of years, e.g. 1.8\,yr in V1974\,Cyg \citep{krautteretal96-1}
and 10\,yr in GQ\,Mus \citep{shanleyetal95-1}. Additional support for
turn-off times on the order of years comes from monitoring of the
ultraviolet luminosity of a number of post novae following their
eruption \citep{gonzales-riestraetal98-1}.

\citet{prialnik86-1} modelled the evolution of a classical nova
through a complete cycle of accretion, outburst, mass loss, decline
and resumed accretion. During the decline, the cooling of the white
dwarf can be fit with a power-law of the form
\begin{equation}
L \propto t^{-1.14},
\end{equation}
where $L$ denotes the luminosity of the white dwarf and $t$ the time
after the nova explosion.  Prialnik's theoretical model is confirmed
by \citet{somers+naylor99-1}, who derive the cooling rate of the white
dwarf using B\,band observations of the irradiated secondary in
V1500\,Cygni and find
\begin{equation}
L \propto t^{-0.94\,\pm\,0.09}.
\end{equation}

\section{Irradiation of the disc by the white dwarf}
External irradiation of the disc by the hot central source can
suppress the thermal instabilities particularly in the inner disc
regions.  In the context of soft X-ray transients,
\citet{tuchmanetal90-1} and \citet{mineshigeetal90-1} calculate the
thermal equilibrium structure of externally irradiated accretion disc
annuli assuming that the irradiation flux is thermalized in the
photosphere of the disc. \citet{king97-1} explains the UV-delay in
dwarf novae taking into account irradiation from the white dwarf which
truncates the inner disc.  \citet{hameuryetal99-1} confirm this claim
only for hot white dwarfs ($T_{\mathrm{wd}}\geq\,40\,000\,\mathrm{K}$)
and find that the depletion of the inner disc must create several
small outbursts between the main outbursts, contrary to observations.
\citet{leachetal99-1} explain the low states of VY\,Scl stars as being
due to the interplay of irradiation from hot white dwarfs and low mass
transfer rates.

For completeness we note that the motivation of \citet{king97-1} and
\citet{hameuryetal99-1} was to account for the ``problem of the UV
delay'' in dwarf novae, although the effects of irradiation of the
inner disc are of general interest for the operation of the accretion
disc limit cycle mechanism. \citet{smak98-1} critically re-examined
the basis for the claim of a problem, and finds that there does not
seem to be one.  Previous studies which had claimed there to be a
problem had utilised various simplifications in their time dependent
models which turned out to be critical - such as the neglect of a
variable outer boundary.  The contraction of the accretion disc during
quiescence has the effect of increasing the local surface density in
the outer disc, and thereby promoting disc instabilities which begin
at large radii.  These so-called ``outside-in''
\citep{cannizzoetal86-1} or ``type A'' \citep{smak84-1} outbursts in
which a narrow spike of enhanced surface density propagates from the
initial site of the instability to the inner edge, are able to produce
outbursts with the observed delay.

In this work we calculate the structure of irradiated accretion discs
in non-magnetic post novae, taking into account the time dependent
flux of the white dwarf cooling down from the hydrogen shell burning
phase, as well as the luminosity of the boundary layer resulting from
accretion onto the white dwarf. The luminosity of the white dwarf is
given by
\begin{equation}
\Lwd(t)=4\pi\,\Rwd^2\,\sigma\,T_{\mathrm{wd}}^4(t)\,\propto\,t^{-1.14},
\end{equation}
where \Rwd\ and \Twd\ denote the radius and the effective temperature
of the white dwarf, respectively, and $\sigma$ is the Stefan-Boltzman
constant.  The boundary layer luminosity is 
\begin{equation}
\Lbl=\alpha_{\mathrm{BL}}\,\frac{G\,\Mwd\,\dot M}{\Rwd}
\end{equation}
where $G$ is the gravitational constant and, as in
\citet{stehle+king99-1} and \citet{leachetal99-1} we take
$\alpha_{\mathrm{BL}}=0.5$.  Assuming that the boundary layer
luminosity is radiated by the entire surface of the white dwarf and
the disc is geometrically thin, the time-dependent flux
$F_\mathrm{irr}(t)$ irradiating the disc at the radius $R$ is given by
\begin{equation}
F_\mathrm{irr}(t)=(1-\beta)
\frac{\Lbl+\Lwd(t)}{2\,\pi\,\sigma\,\Rwd^2}
\frac{1}{\pi}[\arcsin\rho-\rho(1-\rho^2)^{\frac{1}{2}}],
\end{equation}
\citep{adamsetal88-1,king97-1}, where $\beta$ is the albedo,
$\rho=\Rwd/R$ and $t=0$ at the end of the hydrogen shell burning
phase. Throughout this paper we follow \citet{king97-1} in adopting
$\beta=0.5$.

If the irradiation temperature ($\sigma
T_\mathrm{irr}^4\,\equiv\,F_\mathrm{irr}$) exceeds
$T_\mathrm{H}\,\sim\,6500\,\mathrm{K}$ the hydrogen in the disc is
fully ionised independent of the accretion rate
\citep{vanparadijs96-1} Thus, setting
$T_\mathrm{irr}\,\geq\,T_\mathrm{H}$ at the outer edge of the disc
($R=R_\mathrm{out}$) gives a limit of the irradiation flux which
suppresses the disc instability.  Using standard equations
\citep{eggleton83-1,franketal92-1} and assuming that the outer disc
radius $R_{\mathrm{out}}$ is 70\,\% of the primary's Roche lobe
radius, we obtain the outer radius of the accretion disc as function
of the orbital Period $P$ and the binary mass ratio $q=M_2/\Mwd$.  We
used $q=0.5$ for all calculations throughout the paper as we are
mainly interested in systems above the period gap.

We can now estimate how long disc instabilities in the accretion discs
of post novae are suppressed due to irradiation, where we assume
either $\dot M=0$ or $\dot M=10^{17}\mathrm{g\,s^{-1}}$, as typical
for dwarf novae above the period gap (for much higher accretion rates,
the disc remains in a stable, hot state anyway).  Fig.\,1 shows the
results for initial temperatures of $\Twd(0)=5\times 10^5 \mathrm{K}$
and $\Twd(0)=3\times 10^5\,\mathrm{K}$. Apparently, disc
instabilities, and, hence, dwarf nova outbursts, should typically be
suppressed for $\sim$5--50\,yr, and up to $\sim$100\,yr if the white
dwarf is strongly heated during the nova eruption.  It is also clear
that the size of the disc is a crucial parameter: the irradiation from
the white dwarf can suppress the instability over the entire disc only
for orbital periods $\Porb\la 20$\,hr. The contribution of the
boundary layer luminosity becomes important only after $\sim5$\,yr at
the earliest.

It is important to note that our results are lower limits on the time
scale on which irradiation from the white dwarf suppresses dwarf nova
outbursts in post novae for the given disc albedo: (a) in order to
produce significant dwarf nova outbursts, the disc instability has to
affect on considerable parts of the outer disc and not only the outer
edge (we focus on this subject in the next section) and (b) the disc
may be flared and therefore intercept more flux from the white dwarf
than described by Eq.(5), which assumes a flat disc.

\section{Dwarf novae among post novae?}
Accretion disc instabilities and their observational consequence,
dwarf nova outbursts, can occur in a non-magnetic cataclysmic variable,
if under the assumption of stationary accretion, the disc at its
outer rim is colder than the
ionisation temperature of hydrogen, $T_{\mathrm{H}}$. As described above, two
factors determine the disc temperature at a given radius: the mass
transfer rate and the irradiation field from the white dwarf.

While most post novae seem to have no outbursts,  
a small number of post novae show repetitive optical brightenings that
are reminiscent of dwarf nova outbursts
\citep{livio89-1,honeycuttetal98-2}.
The post novae for which this outburst behaviour is best documented
are GK\,Per (\,=\,Nova Per 1901), which shows outbursts of
$\sim3$\,mag every $\sim1000$\,d, and V446\,Her (\,=\,Nova Her 1960),
which shows outbursts with an amplitude of $\sim2.5$ mag and a
recurrence time of $\sim23$\,d \citep{honeycuttetal98-1}.

\begin{figure}
\begin{center}
\includegraphics[angle=0,width=7.9cm]{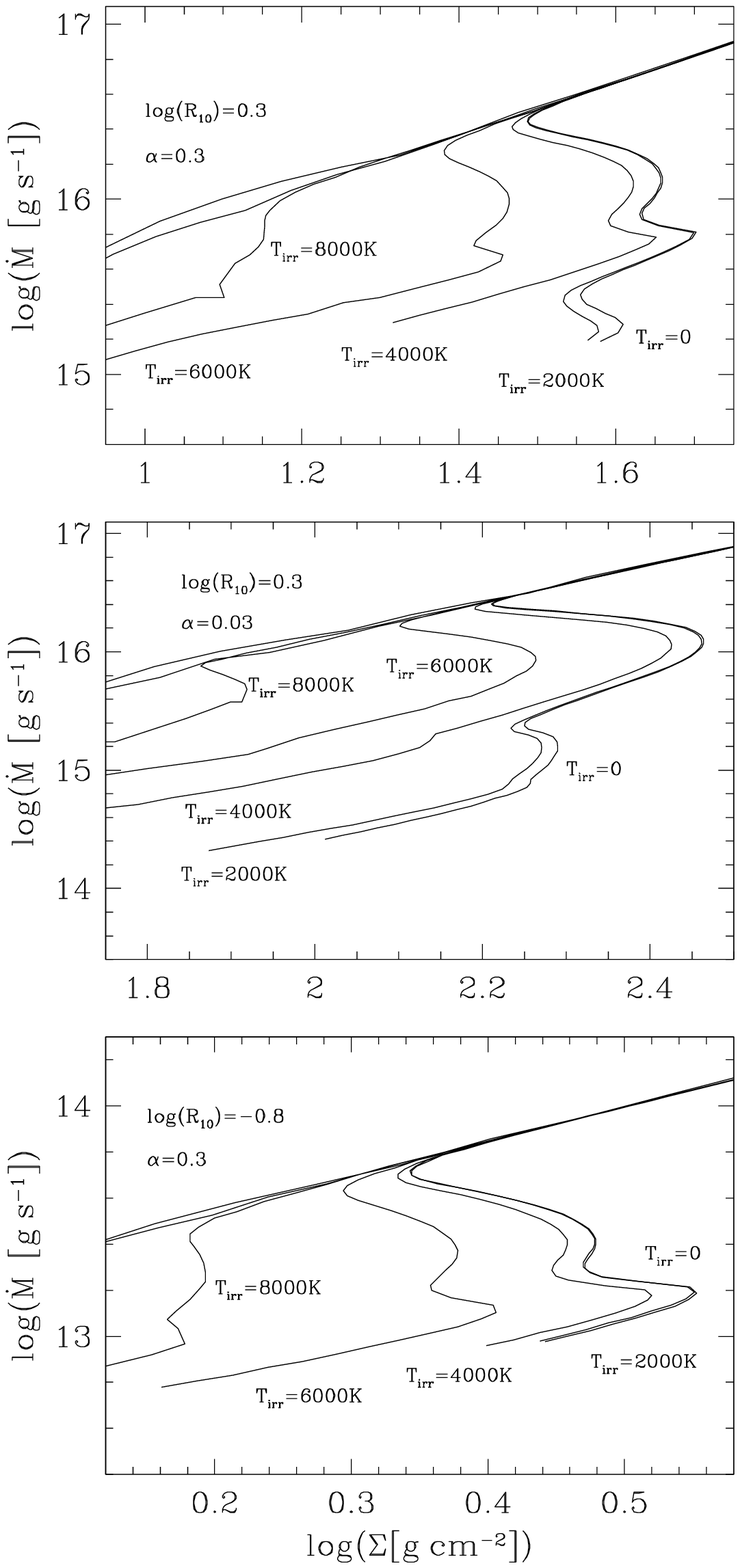}
\caption[]{The vertical equilibrium curves for irradiation
temperatures of $T_{\mathrm{irr}}=[0, 2, 4, 6, 8, 10,
12]\,\times\,10^3\,\mathrm{K}$ (right to left) and a white dwarf mass
of $M_{\mathrm{wd}}=1.3\,M_{\odot}$. 
Assuming a relatively high
$\alpha$ and at a large radius, the disc instability is suppressed for an
irradiation temperature $T_\mathrm{irr}=\,8000$\,K (top panel), 
while it is not
fully supressed for a lower $\alpha$ (middle panel) or at a smaller
radius (bottom panel) and the same irradiation temperature
$T_{\mathrm{irr}}=8000\,\mathrm{K}$.
The external irradiation temperature necessary to suppress the
instability increases with decreasing radius and decreasing $\alpha$ 
(see Eq.\,(8)).
Note the different scalings.}
\end{center}
\end{figure}

The presence of outbursts in GK\,Per is not surprising: The critical
mass transfer rate below which outbursts are possible is ${\dot
M}_{\mathrm{cr}} \simeq 10^{16} $ g s$^{-1}$ $R_{10}^{2.6}
m_1^{-0.87}$, where $R_{10}$ is the outer disc radius in units of
$10^{10}$ cm and $m_1$ is the primary mass in solar units
\citep{cannizzoetal88-1}.  The large disc size ($\Porb=2$\,d) for
GK\,Per means that ${\dot M}_{\mathrm{cr}}\simeq 2\times 10^{19}$
g\,s$^{-1}$, which ensures that disc instabilities are possible even
at very high mass transfer
rates and almost independent of the mass transfer
rate. Indeed, the outbursts observed in GK\,Per have been successfully
modelled within the disc instability scenario
\citep{cannizzo+kenyon86-1,kimetal92-1}.

By contrast, V446\,Her has $\Porb=4.97$\,h
\citep{thorstensen+taylor00-1}, which places it among a number of
``normal'' dwarf novae longward of the $2-3$ hr period gap.  The dwarf
nova nature of V446\,Her is confirmed by its spectrum which is typical
for this class \citep{thorstensen+taylor00-1}.  The presence of dwarf
nova outbursts in V446\,Her indicates that the accretion rate cannot
be too high. 

In the previous section we assumed that the disc instability is
suppressed for $T_{\mathrm{irr}}\geq\,6500\,\mathrm{K}$, which is a
reasonable but somewhat simplified statement. In order to get more
detailed results, we re-derive the accretion rate above which the disc
instability is suppressed, $\dot{M}_{\mathrm{cr}}$, from calculations
of the vertical structure of irradiated accretion discs.  Our
computations are based on the vertical structure code written by one
of us (JKC) and described in \citet{cannizzo+wheeler84-1} and
\citet{cannizzo+cameron88-1} which we have modified to include the
effects of external irradiation.  Assuming that the external flux is
thermalized in the surface layer, we set
\begin{equation}
\sigma\,T_{\mathrm{s}}^4=F_{\mathrm{visc}}+\sigma\,T_{\mathrm{irr}}^4
\end{equation}
at the outer boundary, where $T_{\mathrm{s}}$ is the photospheric
temperature, $F_{\mathrm{visc}}$ the energy flux arising from viscous
dissipation and $T_{\mathrm{irr}}$ the irradiation temperature defined
above (Hameury et al. 1999).  We calculate the modified vertical
structure for $M_{\mathrm{wd}}=\,[0.5,...,1.3]\,M_{\odot},
\alpha=[0.005,...,1],\,R=[0.1,...,10]\times10^{10}\,\mathrm{cm}$ and
$T_{\mathrm{irr}}=[0,...,12]\times\,10^3\,\mathrm{K}$.  For the
non-irradiated disc (i.e. $T_{\mathrm{irr}}=0$), we obtain
\begin{equation}
\dot{M}_{\mathrm{cr}}=9.5\times\,10^{15}\,\mathrm{g\,s^{-1}}\,R_{10}^{2.64}\,\alpha^{0.01}\,m_{\mathrm{1}}^{-0.88}.
\end{equation}
This result is in good agreement with values derived previously
\citep{cannizzoetal88-1,ludwigetal94-1,hameuryetal98-1}.

Fig.\,2 shows examples of the thermal equilibrium curves for different
irradiation temperatures.  Irradiation has a significant influence on
the ionisation state of the disc for
$\log(T_{\mathrm{irr}})\geq\,3.5$.  The critical accretion rate, below
which the disc becomes unstable, decreases with increasing irradiation
until the external flux becomes strong enough to hold the disc in the
hot stable state independent of the accretion rate.  
We find that the irradiation temperature $T_\mathrm{irr,s}$ for which
$\dot{M}_{\mathrm{cr}}$ vanishes is a slightly decreasing function of
$R_{10}$ and $\alpha$, but nearly independent of $m_1$:
\begin{equation}
T_{\mathrm{irr,s}}=7382\,\mathrm{K}\,\alpha^{-0.07}\,R_{10}^{-0.03}.
\end{equation}
The decrease of $\dot{M}_{\mathrm{cr}}$ between the non-irradiated case
(Eq.\,7) and the suppression of the instability is approximately given by:
\begin{equation}
\dot{M}_{\mathrm{cr}}^{\mathrm{irr}}=\dot{M}_{\mathrm{cr}}\left(1-\frac{T_{\mathrm{irr}}^{7.2}}{T_{\mathrm{irr,s}}^{7.2}}\right),
\end{equation}
where $T_{\mathrm{irr,s}}$ and $\dot{M}_{\mathrm{cr}}$ given above.
Using the result of our parameter study, i.e.
$\dot{M}_\mathrm{cr}^{\mathrm{irr}}(M_{\mathrm{wd}},R,T_{\mathrm{irr}},\alpha)$, combined with Eq.\,(5) and assuming $\Twd(0)=3\times\,10^5$\,K after
the nova event, we find that the limit cycle instability can operate
now (i.e. 40\,yr after the 1960 nova) only between
$\sim1.2\,\times\,10^{10}$\,cm and
$R_\mathrm{out}\simeq\,3.8\,\times\,10^{10}$\,cm in V446\,Her
(Fig.\,3).  This might be enough to produce significant dwarf nova
outbursts but the continuing cooling of the white dwarf should lead in
the future to an increase of the outburst amplitude as well as to
longer quiescence periods.

Assuming that the mass transfer rate in V446\,Her is indeed low enough
to permit a disc instability limit cycle, we compare now our estimates
of irradiation suppression of dwarf nova outbursts with the long-term
photometric history of the system.

The pre-nova showed brightness fluctuations between $V\sim18-15$,
which bear some resemblance to dwarf nova outbursts, even though the
rise time of the single well sampled ``flare'' seems too slow compared
to a typical dwarf nova outburst
\citep{robinson75-1}. \citet{stienon71-1} reports 9 photometric
measurements of the post nova obtained between 22 September 1968, and
26 September 1970, i.e. 8--10\,yr after the nova eruption, all of which
show the system at $m_\mathrm{pg}\sim15.8$. Finally, the systematic
RoboScope photometry of \citet{honeycuttetal98-2}, starting in 1990,
shows V446\,Her at a quiescent magnitude of $V\sim18$ with regular
outbursts reaching $V\sim15.5$.  It appears very unlikely that all of
Stienon's measurements caught the system at the maximum of an
outburst, and we believe that dwarf nova outbursts were suppressed in
V446\,Her for at least ten years after the nova eruption in
1960. Unfortunately, we do not know how long the hydrogen shell
burning lasted in V446\,Her, and it is not clear at which point
between 1970 and 1990 the dwarf nova outbursts resumed. Therefore, we
can not give a quantitative estimate of $\Twd(0)$.

\begin{figure}
\includegraphics[angle=0,width=9cm]{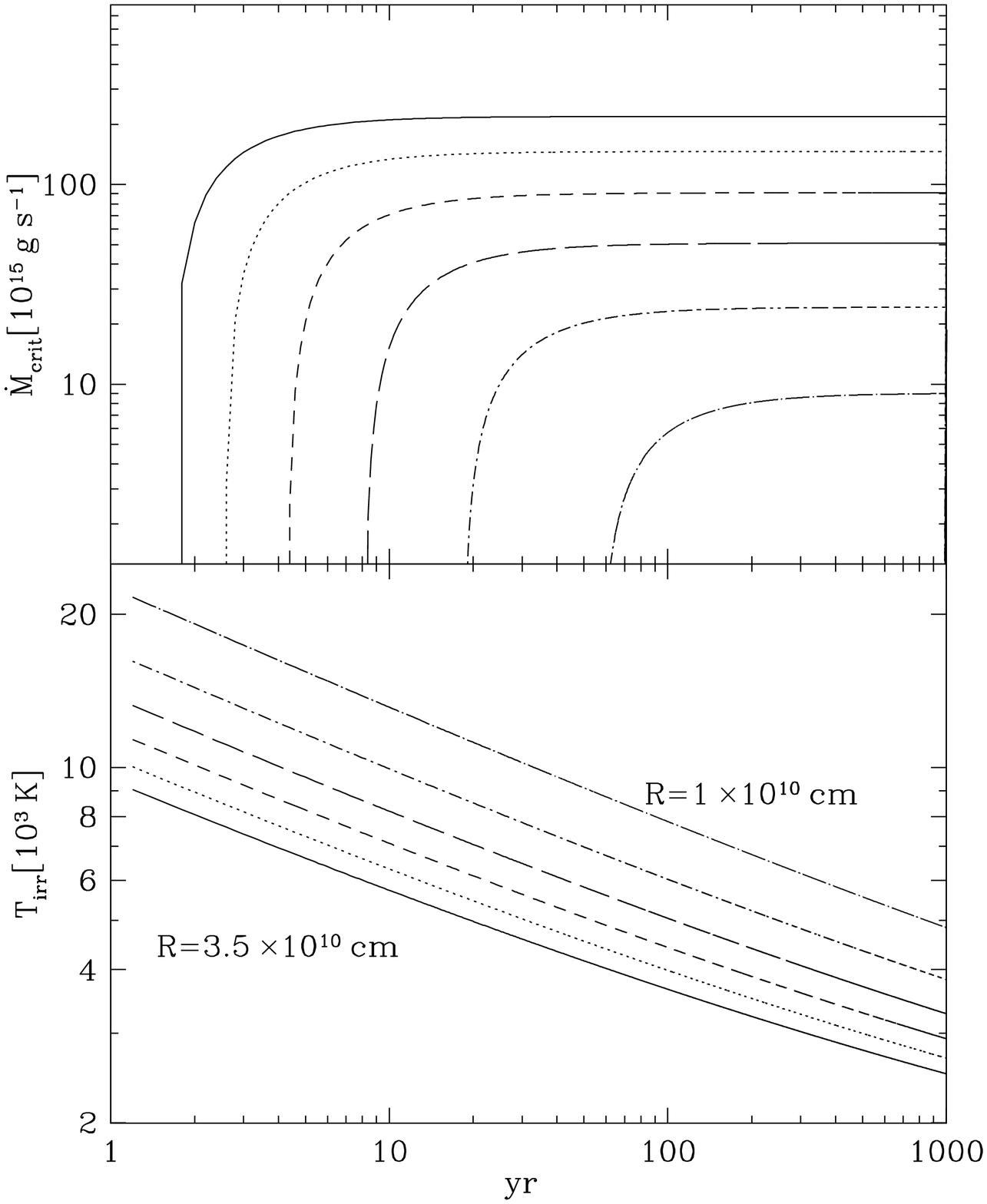}
\caption[]{\label{exam-3} The critical accretion rate and the
irradiation temperature as a function of time for different disc annuli
($R=[3.5,3,2.5,2,1.5,1]\times\,10^{10}\,\mathrm{cm}$), $\alpha=0.3$ and $m_1=0.6$.}
\end{figure}

\section{Discussion}
Taken at face value, the decrease of the optical brightness observed
in V446\,Her over several decades following the nova eruption and the
onset of dwarf nova outbursts $\sim1-3$ decades after the nova
eruption could be interpreted simply as a decline of the accretion
rate, being higher than the critical accretion rate for at least the
first decade after the eruption, and being below the critical rate in
the 1990's.

Such a decline of the accretion rate in post novae is the hallmark of
the ``hibernation scenario'', which was invoked by Shara et
al. (1986\nocite{sharaetal86-1}, see \citealt{shara89-1} for
a review) to solve discrepancies between nova theory and observations
and to account for the low observed space density of CVs when compared
to the number of nova eruptions in our Galaxy or M31.
In the hibernation scenario, irradiation of the secondary star by the
hot white dwarf should keep the mass transfer rate high after the nova
eruption for a limited period of time, after which the mass transfer
rate decreases to very low values. The post novae should, hence,
appear at first as a novalike variables (for many decades to a
century) with stable hot accretion discs, and evolve thereafter into
inconspicuous low $\dot M$ CVs. Thus, in the hibernation scenario it
appears plausible that most ``fresh'' post novae have indeed high
accretion rates, and that the oldest recovered novae are intrinsically
very faint \cite[e.g][]{sharaetal85-1}.

On observational grounds, the hibernation scenario has been
challenged, e.g. \citet{nayloretal92-1} and \citet{weightetal94-1},
but obtained also some support: \citet{vogt90-1} and
\citet{duerbeck92-1} found a gradual decrease in the visual brightness
during the final decline of novae and interpret this result as an
evidence for a decrease of the mass transfer rate in these systems,
consistent with the hibernation scenario. However, the latter point
needs to be considered with some care: as already mentioned in the
introduction, irradiation by the hot white dwarf significantly
contributes to the observed disc brightness. As the white dwarf cools
from the nova outburst, the decreasing amount of irradiation will
result in a decrease of the observed disc brightness, and, hence,
mimick a decrease of the accretion rate.

A valuable alternative to the hibernation model that can explain at
least the fact that only in a small number of cases dwarf nova
outbursts were observed in post novae (and pre novae) is the mass 
transfer cycle
proposed by \citet{kingetal95-1} in which the accretion-induced
irradiation of the secondary drives a limit cycle with a period of
$\sim10^6-10^7$\,yr. In this cycle, a CV spends similar times in
states of high and low accretion rates, appearing as a novalike
variable or as a dwarf nova. It is then clear that the probability
that a CV turns into a nova is highest during the phase of high mass
transfer, which naturally explains why most post novae are novalike
variables. Of course, in some low $\dot M$ CVs the accreted envelope
will also reach the critical mass for a nova eruption, and V446\,Her
might be such a case.  

In this scenario,
pre novae and post novae would have the same characteristics, which is
in agreement with the analysis of \citet{robinson75-1}, who found that
in most novae the pre-eruption and post eruption magnitudes are very
similar.

\section{Conclusion}
We have calculated the effect that irradiation by the heated white
dwarf in a post nova has on the structure of an accretion disc. We
find that, even if the accretion rate in such a system is low enough
to permit disc instabilities, irradiation from the white dwarf
suppresses dwarf nova outbursts for up to  $\sim\,100$\,yr.  In
the case of V446\,Her our calculations predict an increase of the
outburst amplitude and a decrease of the outburst frequency as the
white dwarf keeps on cooling down. We encourage long-term monitoring
of V446\,Her in order to detect any evolution of the outburst pattern.

\begin{acknowledgement}
We thank the referee H. Ritter and F.V. Hessmann for their
helpful comments and suggestions.
MRS would like to thank the Deutsche Forschungsgemeinschaft for
financial support (Ma\,1545\,2-1). BTG thanks for support from the DLR
under grant 50\,OR\,99\,03\,6.
\end{acknowledgement}


\begin{thebibliography}{34}
\expandafter\ifx\csname natexlab\endcsname\relax\def\natexlab#1{#1}\fi

\bibitem[\protect\astroncite{{Adams} et~al.}{1988}]{adamsetal88-1}
{Adams} F.C., {Lada} C. J., {Shu} F. H. 1988, ApJ, 326, 865
 

\bibitem[\protect\astroncite{{Cannizzo}}{1993}]{cannizzo93-1}
{Cannizzo} J.K., 1993, {\em The Limit Cycle Instability in Dwarf Nova Accretion
  Disks\/}, pp. 6--40,   Advanced Series in Astrophysics and Cosmology~9
  (Singapore: World Scientific)

\bibitem[\protect\astroncite{{Cannizzo} \& {Cameron}}{1988}]{cannizzo+cameron88-1}{}
{Cannizzo} J.K., {Cameron} A.G.W., 1988, ApJ 330, 327

\bibitem[\protect\astroncite{{Cannizzo} \&
  {Kenyon}}{1986}]{cannizzo+kenyon86-1}
{Cannizzo} J.K., {Kenyon} S.J., 1986, ApJ Lett. 309, L43

\bibitem[\protect\astroncite{{Cannizzo} \& {Wheeler}}{1984}]{cannizzo+wheeler84-1}{}
{Cannizzo} J.K., {Wheeler} J.C., 1984, ApJS 55, 367 

\bibitem[\protect\astroncite{{Cannizzo} et~al.}{1986}]{cannizzoetal86-1}{}
{Cannizzo} J.K., {Wheeler} J.C., {Polidan} R.S., 1986, ApJ 301, 634 

\bibitem[\protect\astroncite{{Cannizzo} et~al.} {1988}]{cannizzoetal88-1}{}
{Cannizzo} J.K., {Shafter} A.W., {Wheeler} J.C., 1988, ApJ 333, 227 

\bibitem[\protect\astroncite{{Duerbeck}}{1987}]{duerbeck87-1}
{Duerbeck} H.W., 1987, Space Science Reviews 45, 1

\bibitem[\protect\astroncite{{Duerbeck}}{1992}]{duerbeck92-1}
{Duerbeck} H.W., 1992, MNRAS 258, 629

\bibitem[\protect\astroncite{{Eggleton}}{1983}]{eggleton83-1}
{Eggleton} P.P., 1983, ApJ 268, 368

\bibitem[\protect\astroncite{{Frank} et~al.}{1992}]{franketal92-1}
{Frank} J., {King} A.R., {Raine} D.J., 1992, {\em Accretion power in
  astrophysics\/} (Cambridge: Cambridge University Press)

\bibitem[\protect\astroncite{{Gonz{\'a}lez-Riestra} 
et~al.}{1998}]{gonzales-riestraetal98-1} {Gonz{\'a}lez-Riestra} R.,
{Orio} M., {Gallagher} J., 1998, A\&AS 129, 23

\bibitem[\protect\astroncite{{Hameury} et~al.}{1998}]{hameuryetal98-1}
{Hameury} J.M., {Menou} K., {Dubus} G., {Lasota} J.P.,{Hur\'e} J-M.  1998, MNRAS 298, 1048

\bibitem[\protect\astroncite{{Hameury} et~al.}{1999}]{hameuryetal99-1}
{Hameury} J.M., {Lasota} J.P., {Dubus} G., 1999, MNRAS 303, 39

\bibitem[\protect\astroncite{{Honeycutt}
  et~al.}{1998{\natexlab{a}}}]{honeycuttetal98-2}
{Honeycutt} R.K., {Robertson} J.W., {Turner} G.W., 1998{\natexlab{a}}, AJ 115,
  2527

\bibitem[\protect\astroncite{{Honeycutt}
  et~al.}{1998{\natexlab{b}}}]{honeycuttetal98-1}
{Honeycutt} R.K., {Robertson} J.W., {Turner} G.W., {Henden} A.A.,
  1998{\natexlab{b}}, ApJ 495, 933

\bibitem[\protect\astroncite{{Kim} et~al.}{1992}]{kimetal92-1}
{Kim} S.W., {Wheeler} J.C., {Mineshige} S., 1992, ApJ 384, 269

\bibitem[\protect\astroncite{{King}}{1997}]{king97-1}
{King} A.R., 1997, MNRAS 288, L16

\bibitem[\protect\astroncite{{King} et~al.}{1995}]{kingetal95-1}
{King} A.R., {Frank} J., {Kolb} U., {Ritter} H., 1995, ApJ Lett. 444, L37

\bibitem[\protect\astroncite{{Krautter} et~al.}{1996}]{krautteretal96-1}
{Krautter} J., {\"Ogelman} H., {Starrfield} S., {Wichmann} R., {Pfeffermann}
  E., 1996, ApJ 456, 788

\bibitem[\protect\astroncite{{Leach} et~al.}{1999}]{leachetal99-1}
{Leach} R., {Hessman} F.V., {King} A.R., {Stehle} R., {Mattei} J., 1999, MNRAS
  305, 225

\bibitem[\protect\astroncite{{Livio}}{1989}]{livio89-1}
{Livio} M., 1989, in {\em Physics of Classical Novae\/}, {Cassatella} A.,
  {Viotti} R. (eds.),   LNP 369, pp. 342--350 (Heidelberg: Springer)

\bibitem[\protect\astroncite{{Ludwig} et~al.}{1994}]{ludwigetal94-1}
{Ludwig} K., {Meyer-Hofmeister} E., {Ritter} H., 1994, A\&A 290, 473

\bibitem[\protect\astroncite{{MacDonald}}{1996}]{macdonald96-1}
{MacDonald} J., 1996, in {\em Cataclysmic Variables and Related Objects\/},
  {Evans} A., {Wood} J.H. (eds.),   IAU Coll. 158, pp. 281--287 (Dordrecht:
  Kluwer)

\bibitem[\protect\astroncite{{Mineshige} et~al.}{1990}]{mineshigeetal90-1}
{Mineshige} S., {Tuchman} Y., {Wheeler} J.C., 1990, ApJ 359, 176

\bibitem[\protect\astroncite{{Naylor} et~al.}{1992}]{nayloretal92-1}
{Naylor} T., {Charles} P.A., {Mukai} K., {Evans} A., 1992, MNRAS 258, 449

\bibitem[\protect\astroncite{{Prialnik}}{1986}]{prialnik86-1}
{Prialnik} D., 1986, ApJ 310, 222

\bibitem[\protect\astroncite{Robinson}{1975}]{robinson75-1}
Robinson E.L., 1975, AJ 80, 515

\bibitem[\protect\astroncite{{Shanley} et~al.}{1995}]{shanleyetal95-1}
{Shanley} L., {\"Ogelman} H., {Gallagher} J.S., {Orio} M., {Krautter} J., 1995,
  ApJ Lett. 438, L95

\bibitem[\protect\astroncite{{Shara}}{1989}]{shara89-1}
{Shara} M.M., 1989, PASP 101, 5

\bibitem[\protect\astroncite{{Shara} et~al.}{1985}]{sharaetal85-1}
{Shara} M.M., {Moffat} A.F.J., {Webbink} R.F., 1985, ApJ 294, 271

\bibitem[\protect\astroncite{{Shara} et~al.}{1986}]{sharaetal86-1}
{Shara} M.M., {Livio} M., {Moffat} A.F.J., {Orio} M., 1986, ApJ 311, 163

\bibitem[\protect\astroncite{{Smak}}{1984}]{smak84-1}
{Smak} J.I., 1984, Acta Astron. 34, 161

\bibitem[\protect\astroncite{{Smak}}{1998}]{smak98-1}
{Smak} J.I., 1998, Acta Astron. 48, 677

\bibitem[\protect\astroncite{{Somers} \& {Naylor}}{1999}]{somers+naylor99-1}
{Somers} M.W., {Naylor} T., 1999, A\&A 352, 563

\bibitem[\protect\astroncite{{Starrfield} et~al.}{1998}]{starrfieldetal98-1}
{Starrfield} S., {Truran} J., {Sparks} M., et~al., 1998, in {\em Wild Stars in
  the Old West: Proceedings of the 13th North American Workshop on CVs and
  Related Objects\/}, {Howell} S., {Kuulkers} E., {Woodward} C. (eds.), pp.
  352--367 (ASP Conf. Ser. 137)

\bibitem[\protect\astroncite{{Stehle} \& {King}}{1999}]{stehle+king99-1}
{Stehle} R., {King} A.R., 1999, MNRAS 304, 698

\bibitem[\protect\astroncite{Stienon}{1971}]{stienon71-1}
Stienon F.M., 1971, PASP 83, 363

\bibitem[\protect\astroncite{{Thorstensen} \&
  {Taylor}}{2000}]{thorstensen+taylor00-1}
{Thorstensen} J.R., {Taylor} C., 2000, MNRAS 312, 629

\bibitem[\protect\astroncite{{Tuchman} et~al.}{1990}]{tuchmanetal90-1}
{Tuchman} Y., {Mineshige} S., {Wheeler} J.C., 1990, ApJ 359, 164


\bibitem[\protect\astroncite{{van Paradijs}}{1996}]{vanparadijs96-1}
{van Paradijs} J., 1996,  ApJ,  464, L139


\bibitem[\protect\astroncite{{Vogt}}{1990}]{vogt90-1}
{Vogt} N., 1990, ApJ 356, 609

\bibitem[\protect\astroncite{{Weight} et~al.}{1994}]{weightetal94-1}
{Weight} A., {Evans} A., {Naylor} T., {Wood} J.H., {Bode} M.F., 1994, MNRAS
  266, 761

\end{thebibliography}

\end{document}